\documentclass[a4paper,11pt]{article}
\pdfoutput=1  

\usepackage{jinstpub} 

\usepackage[utf8]{inputenc} 
\usepackage{booktabs}
\usepackage[export]{adjustbox}
\usepackage{graphicx}
\usepackage{tablefootnote}
\usepackage{threeparttable}
\usepackage[caption=false]{subfig}
\usepackage{textpos}

\title{Characterization of the ePix100a and the FastCCD Semiconductor Detectors for the European XFEL}

\author[a,b,1]{I. Kla\v{c}ková,\note{Corresponding author.}}
\author[c]{G. Blaj,}
\author[d]{P. Denes,}
\author[c]{A. Dragone,}
\author[a]{S. G\"{o}de,}
\author[a]{S. Hauf,}
\author[e]{F. Januschek,}
\author[d]{J. Joseph,}
\author[a]{M. Kuster}

\affiliation[a]{European XFEL, Holzkoppel 4, 22869 Schenefeld, Germany}
\affiliation[b]{Faculty of Electrical Engineering and Information Technology, Slovak University of Technology, Ilkovi\v{c}ova 3,
81219 Bratislava, Slovakia}
\affiliation[c]{SLAC National Accelerator Laboratory, Sand Hill Road 2575, Menlo Park, California 94025, U.S.A.}
\affiliation[d]{Lawrence Berkeley National Laboratory, Cyclotron Road 1, Berkeley, California 94720, U.S.A.}
\affiliation[e]{Deutsches Elektronen-Synchrotron DESY, Notkestraße 85, 22607 Hamburg, Germany}

\emailAdd{ivana.klackova@xfel.eu}

\abstract{The European X-ray Free Electron Laser (EuXFEL) is a research facility providing spatially coherent X-ray flashes in the energy range from $0.25\,$keV to $25\,$keV of unprecedented brilliance and with unique time structure: X-ray pulses with a $4.5\,$MHz repetition rate arranged in trains with 2700 pulses every $100\,$ms. The facility operates three photon beamlines called SASE 1, SASE 2 and SASE 3. Each of the beamlines is hosting two scientific experiments. The SASE 1 beamline started its user operation in September 2017, followed by successful first lasing at the SASE 2 beamline in May 2018. Early user experiments are planned to start in 2019 at this beamline, while early user experiments for the SASE 3 beamline are scheduled for the end of 2018. The quality of the experimental data will gain substantial benefits from an accurate characterization and calibration of the X-ray detectors. Supplementing high repetition rate detectors at MHz speeds, slower detectors such as the ePix100a and the FastCCD will be operated at the train repetition rate of $10\,$Hz. These 2D silicon pixelized detectors use fast parallel column-wise readout implemented as a CCD or as a hybrid pixel detector. In the following, characterization and analysis approaches for the FastCCD and the ePix100a detectors are discussed and the performance of the detectors is evaluated using appropriate state-of-the-art analysis techniques.}


\arxivnumber{1812.01390} 

\proceeding{20$^{\text{th}}$ International Workshop on Radiation Imaging Detectors\\
  24--28 June 2018\\ Sundsvall, Sweden}

\begin{document}

\begin{textblock*}{3cm}(11.5cm,-2cm)
   \fbox{\footnotesize SLAC-PUB-17361}
\end{textblock*}

\setlength{\abovedisplayskip}{1pt}
\setlength{\belowdisplayskip}{1pt}
\setlength{\abovedisplayshortskip}{1pt}
\setlength{\belowdisplayshortskip}{1pt}
\setlength{\textfloatsep}{10pt plus 1.0pt minus 2.0pt}

\maketitle
\flushbottom

\section{Introduction}
\label{sec:intro}

The European X-ray Free Electron Laser (EuXFEL) is an international
research facility in the Hamburg/Schenefeld area, Germany with the
purpose of enabling scientific experiments in various areas using
ultrafast spatially coherent X-ray flashes with unprecedented
brilliance. The facility provides X-rays in the energy range from
$0.25\,$keV to $25\,$keV with a unique time structure: pulses with a
$4.5\,$MHz repetition rate are arranged in trains of $2700$ pulses,
repeating every $100\,$ms, the pulse duration is less than $100\,$fs
\cite{altarelli2011european}. EuXFEL serves three scientific beamlines, each
hosting two scientific instruments, namely:

\begin{itemize}
\item At \emph{SASE 1} the Single Particles, Clusters, and Biomolecules \& Serial Femtosecond Crystallography (\emph{SPB/SFX}) and Femtoscond X-ray experiments (\emph{FXE}) instruments \cite{tschentscher2017photon}.  
\item At \emph{SASE 2} the Materials Imaging and Dynamics (\emph{MID}) and High Energy Density matter (\emph{HED}) instruments \cite{tschentscher2017photon}.
\item At \emph{SASE 3} the Spectroscopy \& Coherent Scattering (\emph{SCS}) and Small Quantum Systems (\emph{SQS}) instruments \cite{tschentscher2017photon}.
\end{itemize}

\subsection{Detectors at EuXFEL}
Requirements deriving from time structure, intensity of the FEL beam
and each individual beamline mandate for novel detectors being
developed as no existing detector technology was capable of simultaneously satisfying
all the requirements~\cite{graafsma2009requirements}. To match the time structure of
EuXFEL photon beam, detectors recording images at 4.5$\,$MHz,
resolving individual pulses and with ability to detect extremely short
pulses (<100$\,$fs) were needed. High dynamic range enabling single
photon counting and simultaneous detection of up to
$10^6\,$ph/pixel/pulse, spatial resolution required by scientific
instrument, energy range matching instrument beamlines and data
processing capabilities of the order of $10-12\,$GB/s per 4.5$\,$MHz
detector are other examples of demanding requirements for detectors at
EuXFEL. A summary of the 2D imaging detectors presently in use or
planned to be used in the near future at the European XFEL is listed
in Table~\ref{tab:detectors}.

\begin{table}[t]
\centering
\caption{\label{tab:detectors} Performance parameters of the 2D imaging pixel detectors in use at the European XFEL.}
\smallskip
\begin{threeparttable}
\begin{tabular}{@{}|l|ccccc|@{}}\hline
\multicolumn{1}{|c|}{2D detector} & \begin{tabular}[c]{@{}c@{}}Pixel size \\ ($\mathrm{\mu m^2}$)\end{tabular} & \begin{tabular}[c]{@{}c@{}}Energy \\ (keV)\end{tabular} & Dynamic Range                      & Frame Rate & Instrument \\ \hline
AGIPD \cite{agipd}      & \begin{tabular}[c]{@{}c@{}}200$\times$200\end{tabular}    & $3-16$     & 10$^4\,$ph @ 12$\,$keV       & 4.5$\,$MHz & SPB, MID   \\
DSSC \cite{dssc}        & \begin{tabular}[c]{@{}c@{}}204$\times$204\tnote{*}\end{tabular}  & $0.5-6$    & 10$^4\,$ph @ 1$\,$keV        & 4.5$\,$MHz & SCS, SQS   \\
ePix100a \cite{epix, blaj2016x}    & \begin{tabular}[c]{@{}c@{}}50$\times$50\end{tabular}      & $3-20$     & 100 ph @ 8$\,$keV           & 240$\,$Hz  & HED, MID   \\
ePix10Ka \cite{blaj2018performance}    & \begin{tabular}[c]{@{}c@{}}100$\times$100\end{tabular}      & $3-25$     & 10$^4\,$ph @ 8$\,$keV           & 240$\,$Hz  & HED  \\
FastCCD \cite{fastccd, januschek2016}  & \begin{tabular}[c]{@{}c@{}}30$\times$30 \\ \end{tabular}      & $0.25-6$   & 10$^3\,$ph @ 0.5$\,$keV      & 120$\,$Hz  & SCS        \\
JUNGFRAU \cite{jungfrau}& \begin{tabular}[c]{@{}c@{}}75$\times$75 \\ \end{tabular}      & $3-25$     & 10$^4\,$ph @ 12$\,$keV       & 1.1$\,$kHz & HED        \\
LPD \cite{lpd}          & \begin{tabular}[c]{@{}c@{}}500$\times$500 \\ \end{tabular}    & $5-20$     & 10$^5\,$ph @ 12$\,$keV       & 4.5$\,$MHz & FXE        \\
pnCCD \cite{pnccd} & \begin{tabular}[c]{@{}c@{}}75$\times$75 \\ \end{tabular} & $0.03-25$ & 6$\times$10$^3\,$ph @ 1$\,$keV & 150$\,$Hz & SCS, SQS \\ \hline
\end{tabular}
\begin{tablenotes}
\item[*] Hexagonal pixels.
\end{tablenotes}
\end{threeparttable}
\end{table}

Exemplary technical requirements for $10\,$Hz detection of the three scientific
instruments \emph{HED}, \emph{MID} and \emph{SCS} are shown in
Table~\ref{tab:BeamReq}. The FastCCD and the ePix100a fulfil these
requirements and have been selected as suitable detectors for the
mentioned scientific instruments.

\begin{table}[b]
\centering
\caption{\label{tab:BeamReq} Technical requirements of the scientific instruments HED, MID and SCS for $10\,$Hz applications detectors.}
\smallskip
\begin{tabular}{@{}|lr|lr|@{}}
\hline
Requirements      & Actual parameters                         & Requirements     & Actual parameters             \\ \hline
\emph{HED \& MID} & \emph{ePix100a}                    & \emph{SCS}       & \emph{FastCCD}         \\ \hline
Compact design    & $(52\times52\times155)\,$mm        & Passage for beam & Beamhole in the sensor \\
Small pixel size  & $50\times 50$\,$\mu m^2$            & Small pixel size & $30\times 30$\,$\mu m^2$\\
Vacuum operation  & Compatible                         & Vacuum operation & Compatible             \\
\begin{tabular}[c]{@{}l@{}}Single photon sensitivity - \\ Low noise \textless{}80 e$^-$ rms\end{tabular} & 41 e$^-$ rms & \begin{tabular}[c]{@{}r@{}}Low noise - \\ \textless{}25 e$^-$ rms\end{tabular} & 23 e$^-$ rms \\
Hard X-ray regime & $3-20\,$keV range & Soft X-ray regime & $0.25-6$\,keV range                   \\ \hline
\end{tabular}
\end{table}

\section{Calibration of the ePix100a and the FastCCD Detectors}
\label{sec:cal}

\begin{figure}[h]
\centering 
\includegraphics[width=.33\textwidth, trim={0 0.25cm 0 0.2cm},clip]{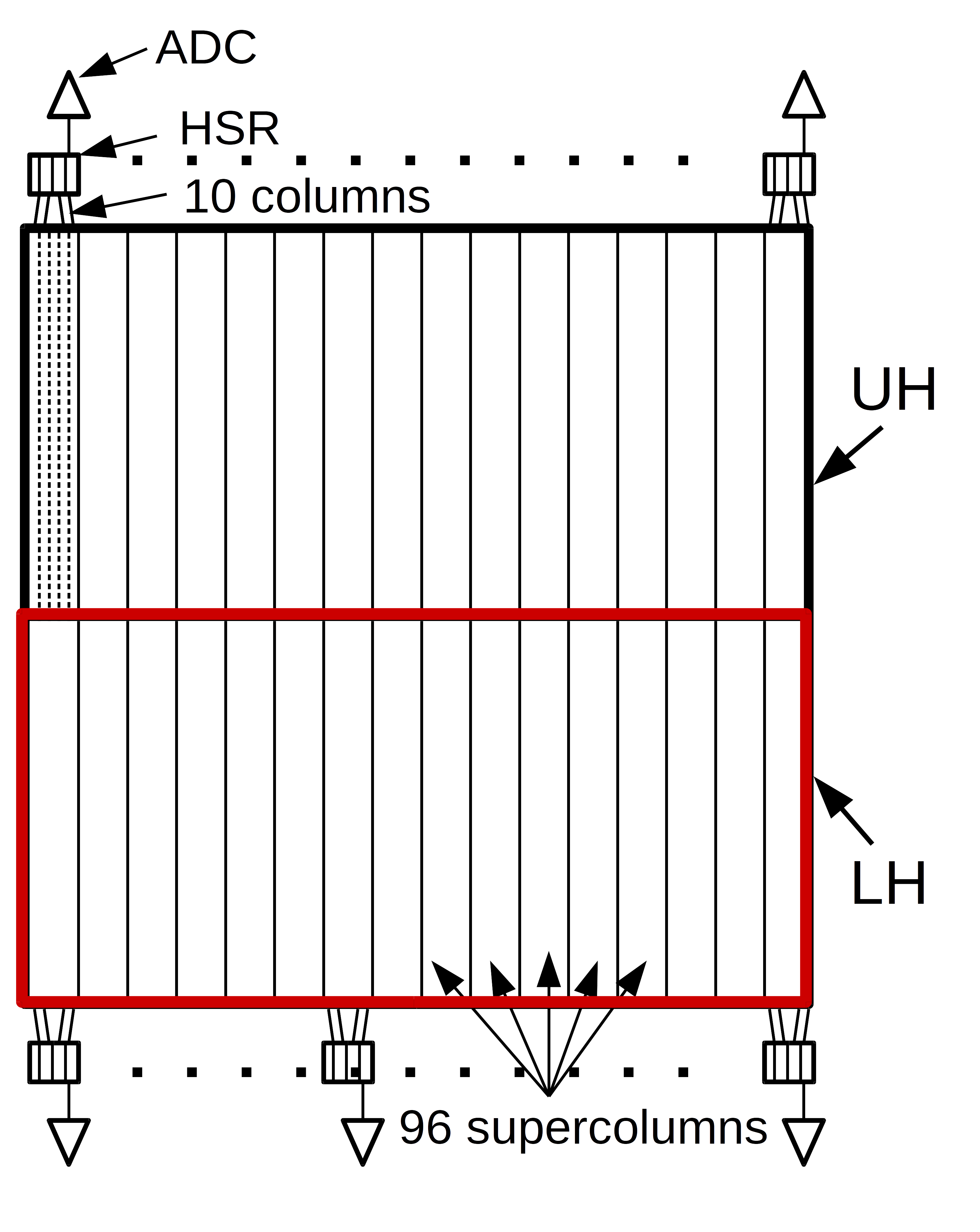}
\qquad
\includegraphics[width=.33\textwidth, trim={0 0.25cm 0 0.2cm},clip]{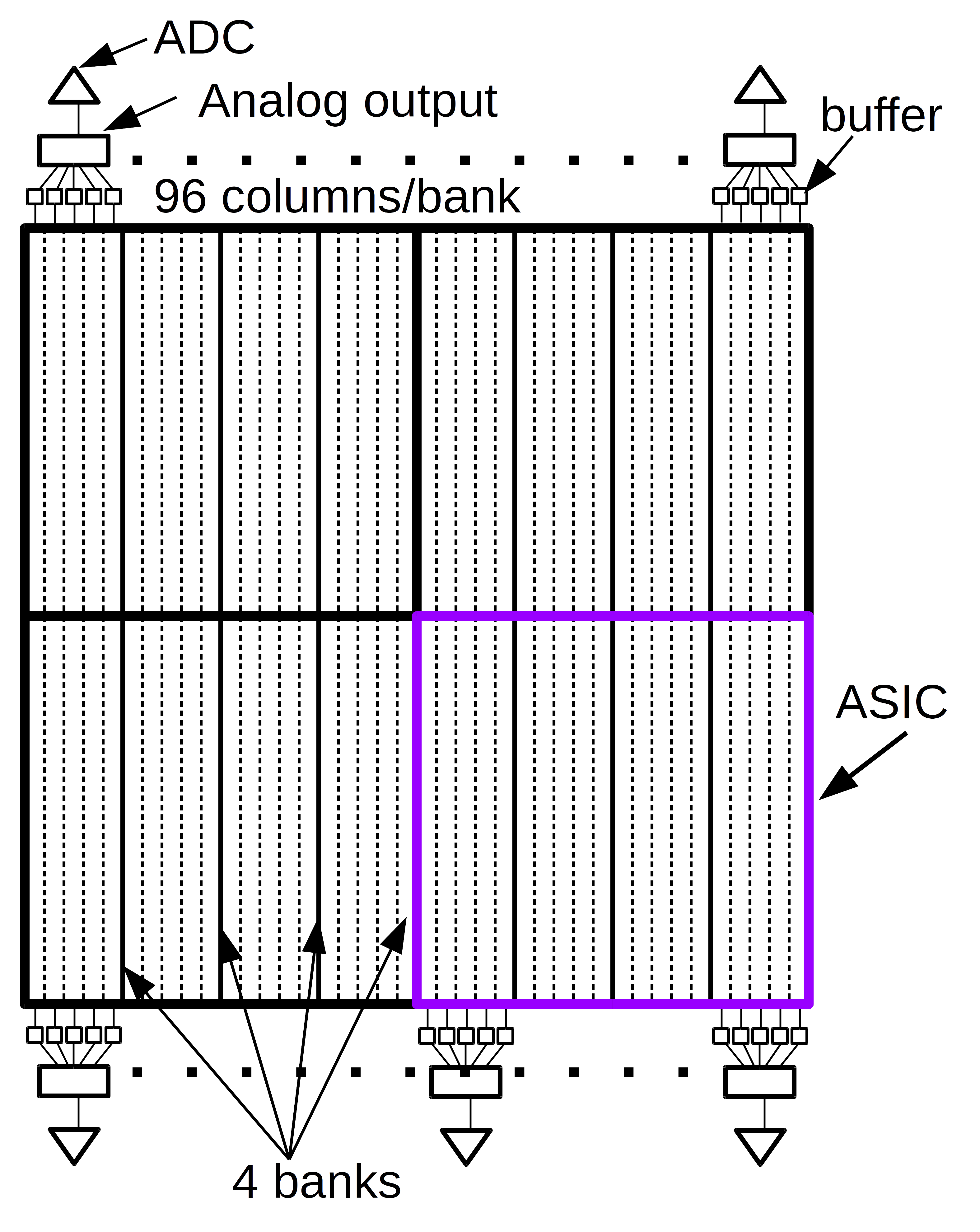}
\caption{\label{fig:readout}Left: Schematic drawing of the FastCCD readout. Ten columns each are compressed to a horizontal shift register (HSR), forming supercolumns. Right: Diagram of the intermediate readout design of the ePix100a. The ASICs are divided into four banks and the columns of each bank are multiplexed to a single analog output, which is digitized by an external ADC.}
\end{figure}

Even though the ePix100a, as a hybrid pixel detector, and the FastCCD,
as a CCD, are following two fundamentally different detector designs, each requiring different calibration method,
similarities exist between the two, as both detectors are read out in
a (quasi-)column-parallel manner. The FastCCD as a CCD detector
performs column-wise charge shifting in order to read out the signal
from pixels. Readout is implemented in a so-called
quasi-column-parallel fashion, where 10 columns are multiplexed onto
one common readout node as shown in the left image of
Figure~\ref{fig:readout}. In addition the sensor is split into two
halves, such that the bottom and top half can be read out separately.

Charge created in the ePix100a pixels is first pre-amplified, integrated
in a single-stage charge integrator, low-pass filtered, baseline 
subtracted and sampled by a column buffer and then digitized in a
column-parallel fashion. The ePix100a consists of four ASICs, each
divided into four banks~\cite{epix}. Columns per-bank are multiplexed
on a single analog output and are digitized by external ADCs, shown
in the right panel of Figure~\ref{fig:readout}. These similarities and differences in
signal handling during read-out allow comparing analysis approaches 
to improve efficiency of calibration and
characterization across different detectors technologies.

\subsection{Calibration Measurements}

Calibration measurements of
the FastCCD were performed using an $^{55}$Fe radiation source, producing Mn K$_{\alpha}$ at 5.9$\,$keV and Mn K$_{\beta}$ peak at 6.49 keV. Measurements were carried out under
high vacuum conditions at $10^{-5}\,$mbar, at a temperature of
$-57\,^{\circ}$C using an exposure time of $50\,$ms. The ePix was operated
under ambient conditions while being irradiated with an X-ray tube
with Mo target and an $^{55}$Fe source. $^{55}$Fe data were acquired
at 17$\,^{\circ}$C with 800$\,\mu$s integration time as cooling of the detector to lower temperatures at the time was limited. 
X-ray tube data were acquired at 10$\,^{\circ}$C with 800$\,\mu$s
integration time.

\section{Performance Evaluation}
\label{sec:eval}
The corresponding data were analysed using the calibration pipeline
developed at EuXFEL for fast, consistent and reliable detector
calibration \cite{kuster2014detectors, XFELcal, XFELdet}. 
The analysis procedures can be divided into two
categories by the type of detector images used. 
The first category of routines uses dark data to estimate effects of the dark signal, e.g., offset and noise. The outcome of the so called "dark calibration" is later needed for the second category of characterisation algorithms, operating on data with X-ray signal, such as flat-field calibration which reveals more detailed information about, e.g., the gain and the energy resolution of the detector.

\subsection{FastCCD Performance}

\begin{figure}[t]
\centering 
\includegraphics[width=.47\textwidth, trim=0 140 0 140, clip=True]{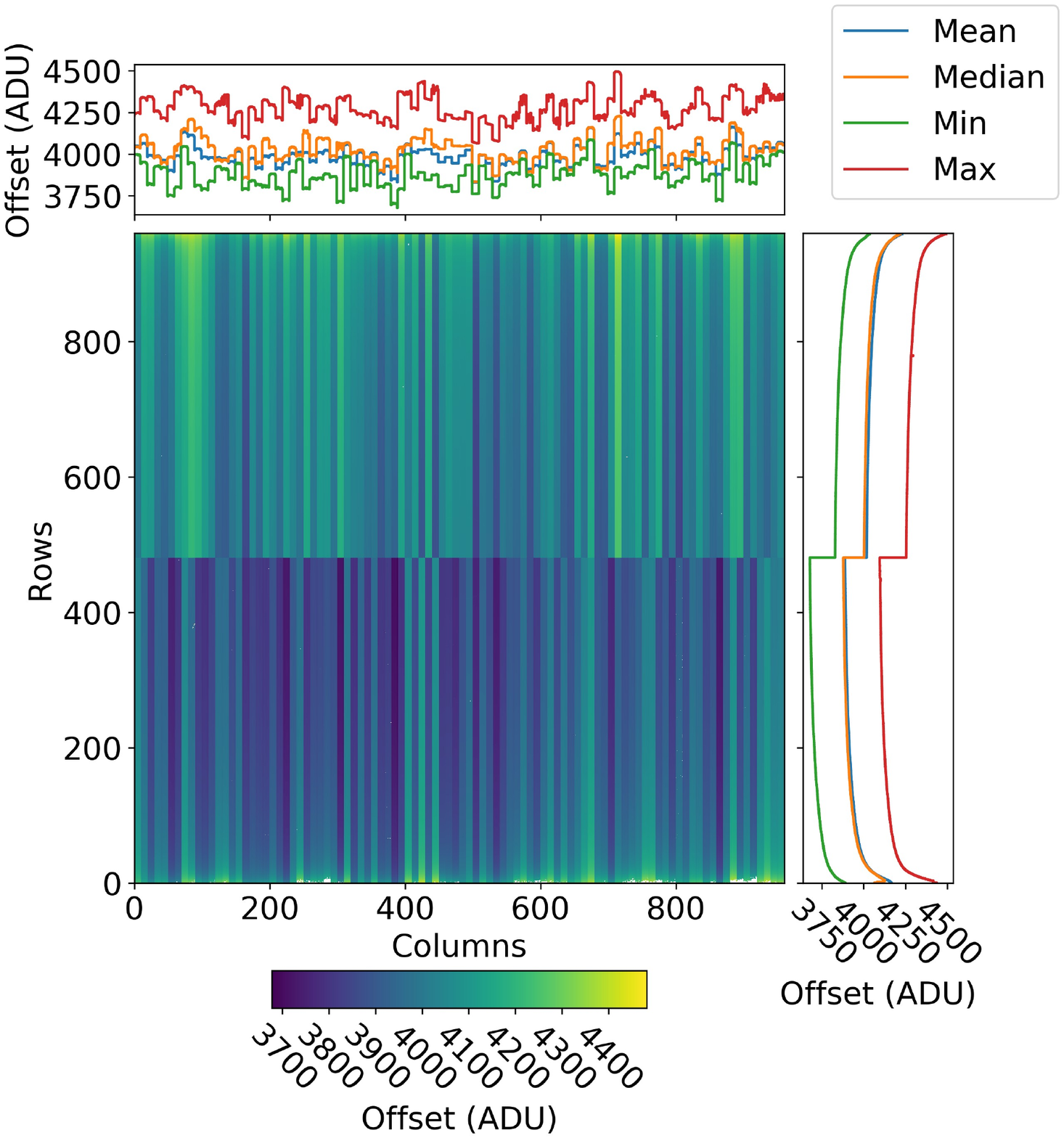}
\qquad
\includegraphics[width=.47\textwidth, trim=0 140 0 140, clip=True]{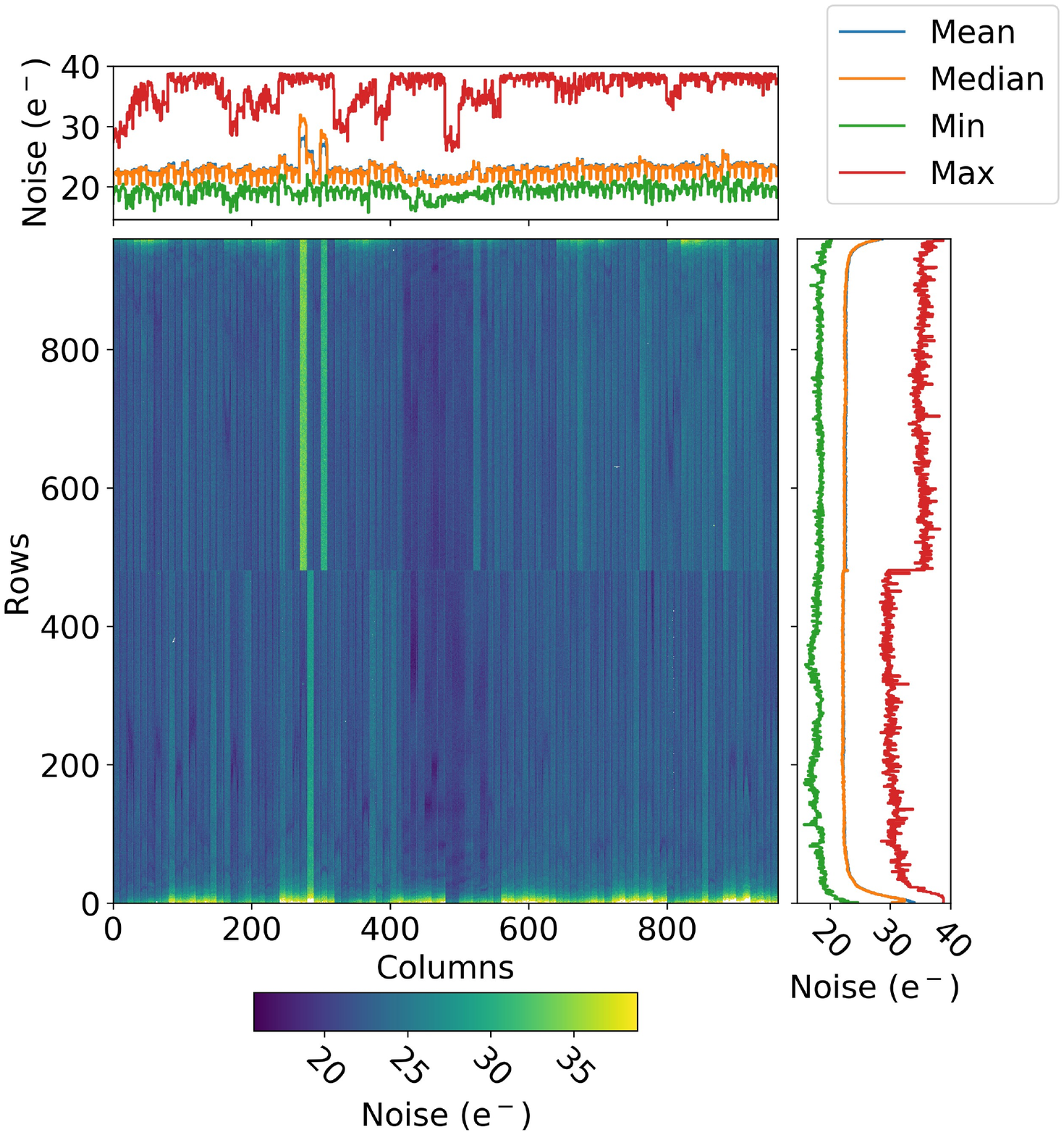}
\caption{\label{fig:darkFCCD} Offset map (left plot, in analog-to-digital units, ADU) and noise map (right plot) of the FastCCD. The side panels represent profiles along the column and row direction.}
\end{figure}

The full size of the FastCCD sensor is $1920\times960$
pixels. However, the camera was operated in the so-called frame store
mode, where only $960\times960$ central pixels are used for charge
integration, the remaining sensor area acts as a fast storage
area, allowing it to operate at higher acquisition rates. 
Evaluating dark images reveals general performance related
features of a detector, e.g. prominent strip pattern observable in an
offset map as shown in Figure~\ref{fig:darkFCCD}. Due to the
quasi-column-parallel readout of the FastCCD, a strip-like pattern
corresponding to so-called supercolumns is apparent in the dark image
(Figure~\ref{fig:darkFCCD}, left).
Note that mean and median values of row and column profiles show similar trend, which is due to the homogeneous response of the sensor across the whole sensitive area, with no significant outliers occurring.
Another feature relates to the sensor design, being divided into two
hemispheres (top and bottom) and read out separately towards the top
and bottom. The two hemispheres are distinguishable in both the
offset and the noise map (Figure~\ref{fig:darkFCCD}, left and right).
As RMS noise our analysis reveals ($22.7\pm2.2$)$\,\mathrm{e}^-$ for
the FastCCD. Noisier edges apparent in Figure~\ref{fig:darkFCCD} are
caused by the readout electronics.

\begin{figure}[tbp]
\centering 
\minipage{0.42\textwidth}%
	\includegraphics[width=\linewidth, valign=t]{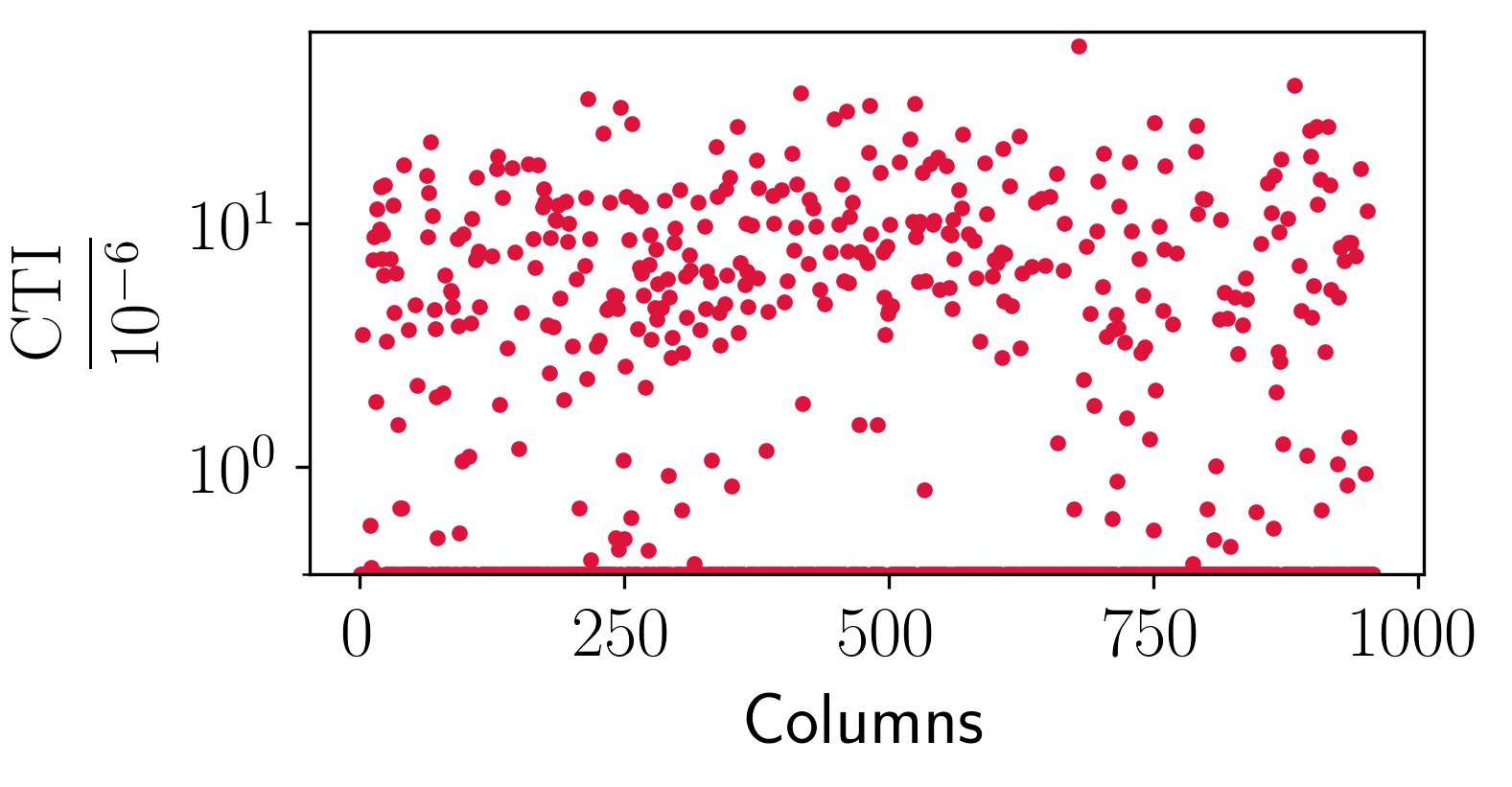}
	\vfill
	\includegraphics[width=\linewidth, valign=t]{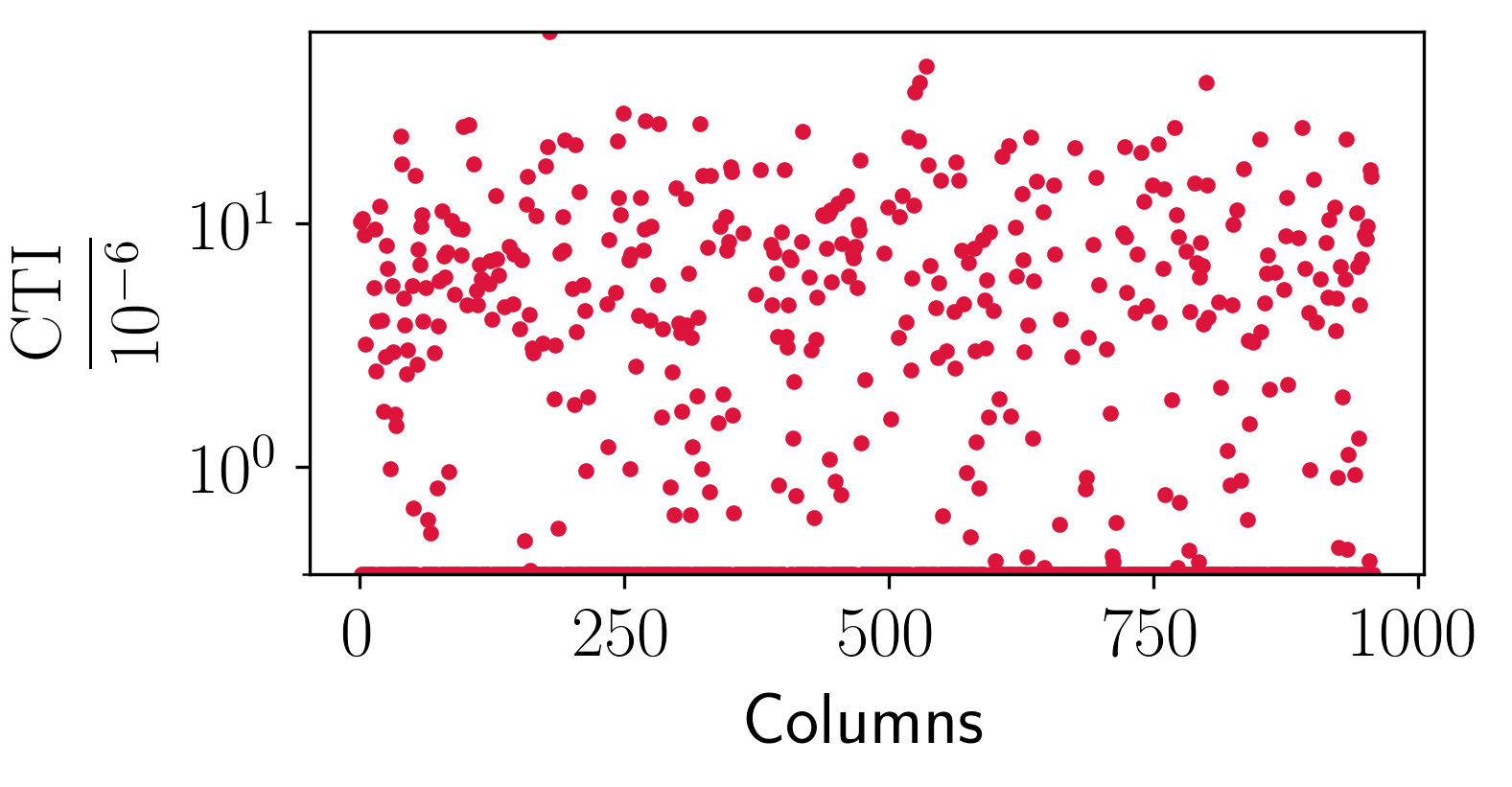}%
	\caption{Column-wise CTI of the FastCCD for the lower (top)
	and upper (bottom) hemisphere}\label{fig:columnCTI}
\endminipage \hfill
\minipage{0.38\textwidth}
	\includegraphics[width=\linewidth, valign=t]{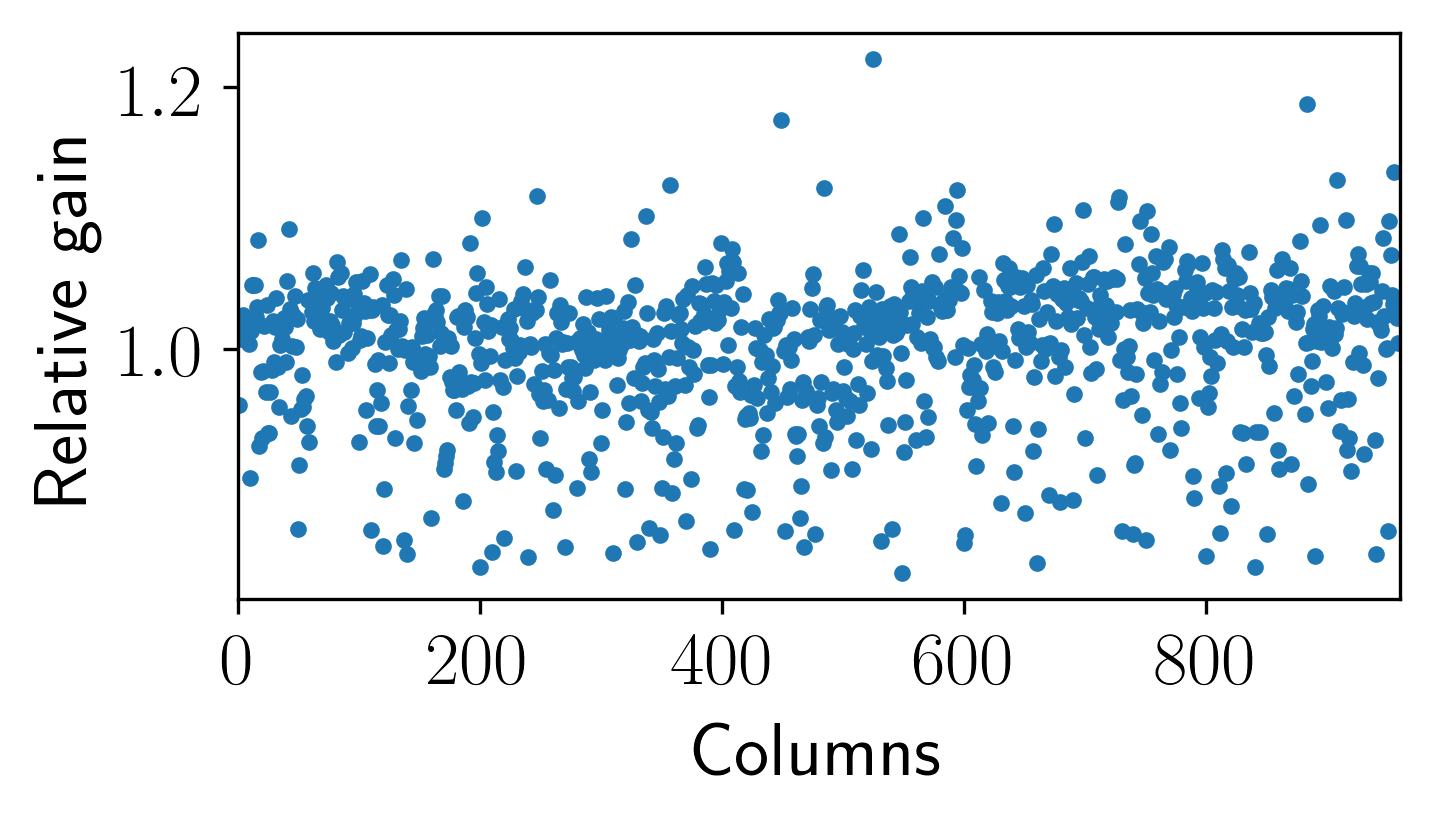}
	\includegraphics[width=\linewidth, valign=t]{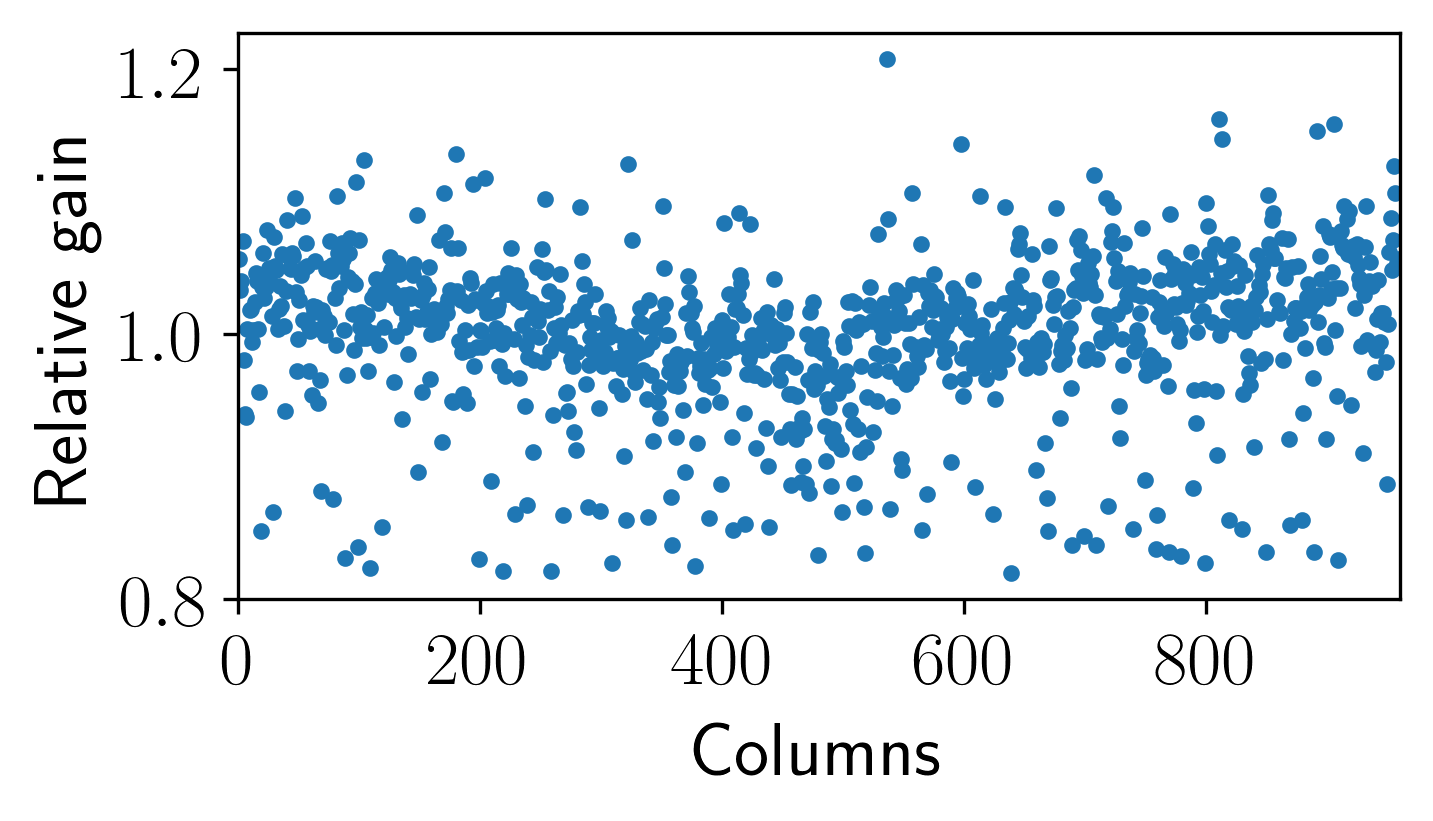}%
	\caption{Column-wise relative gain of the FastCCD for the
	lower (top) and the upper (bottom) hemisphere. }\label{fig:columnRG}
\endminipage
\end{figure}

\begin{figure}[htbp]
\centering 
\includegraphics[width=.47\linewidth, valign=t, trim=0 140 0 140, clip=True]{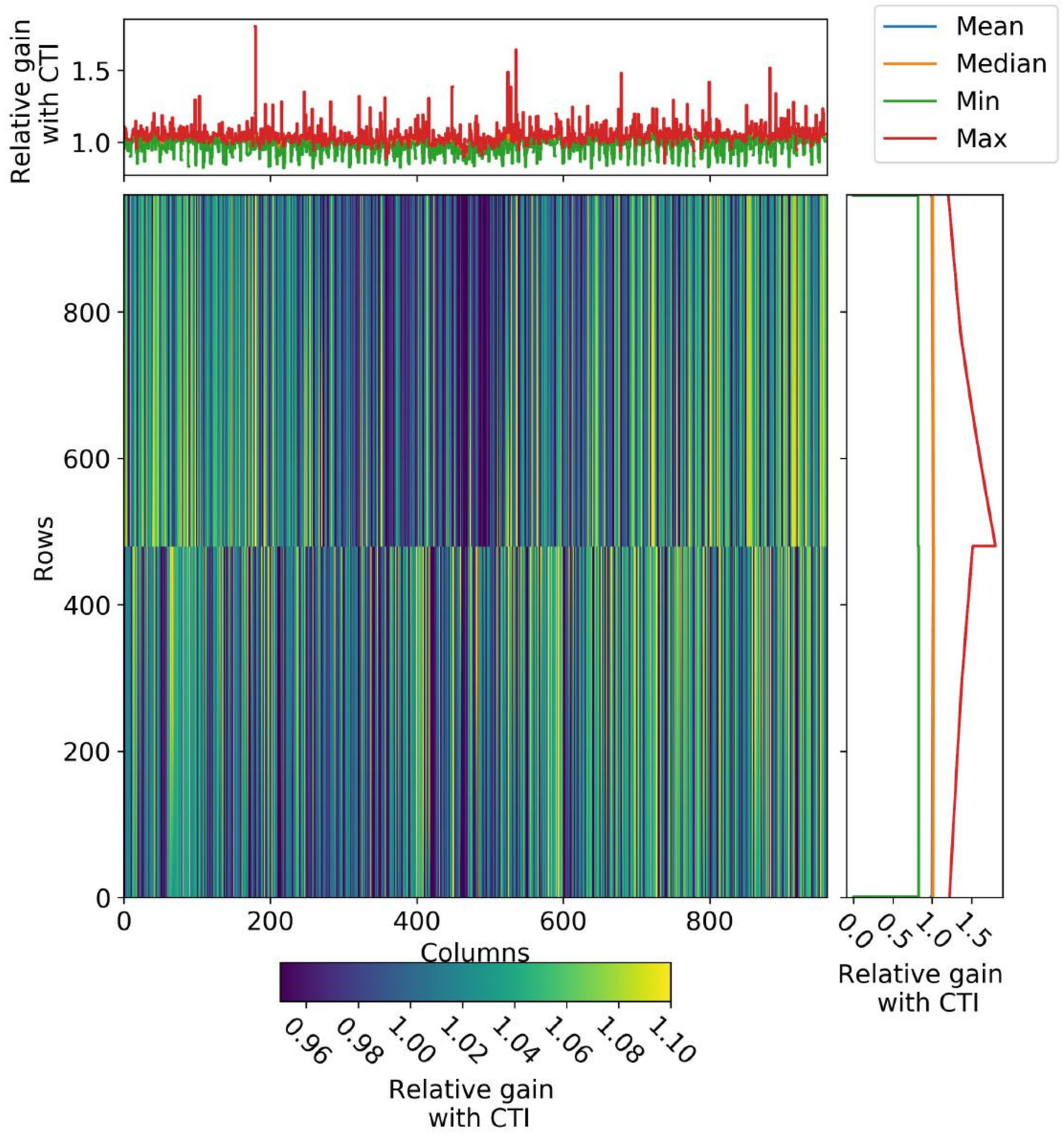}
\qquad
\includegraphics[width=.45\linewidth, valign=t]{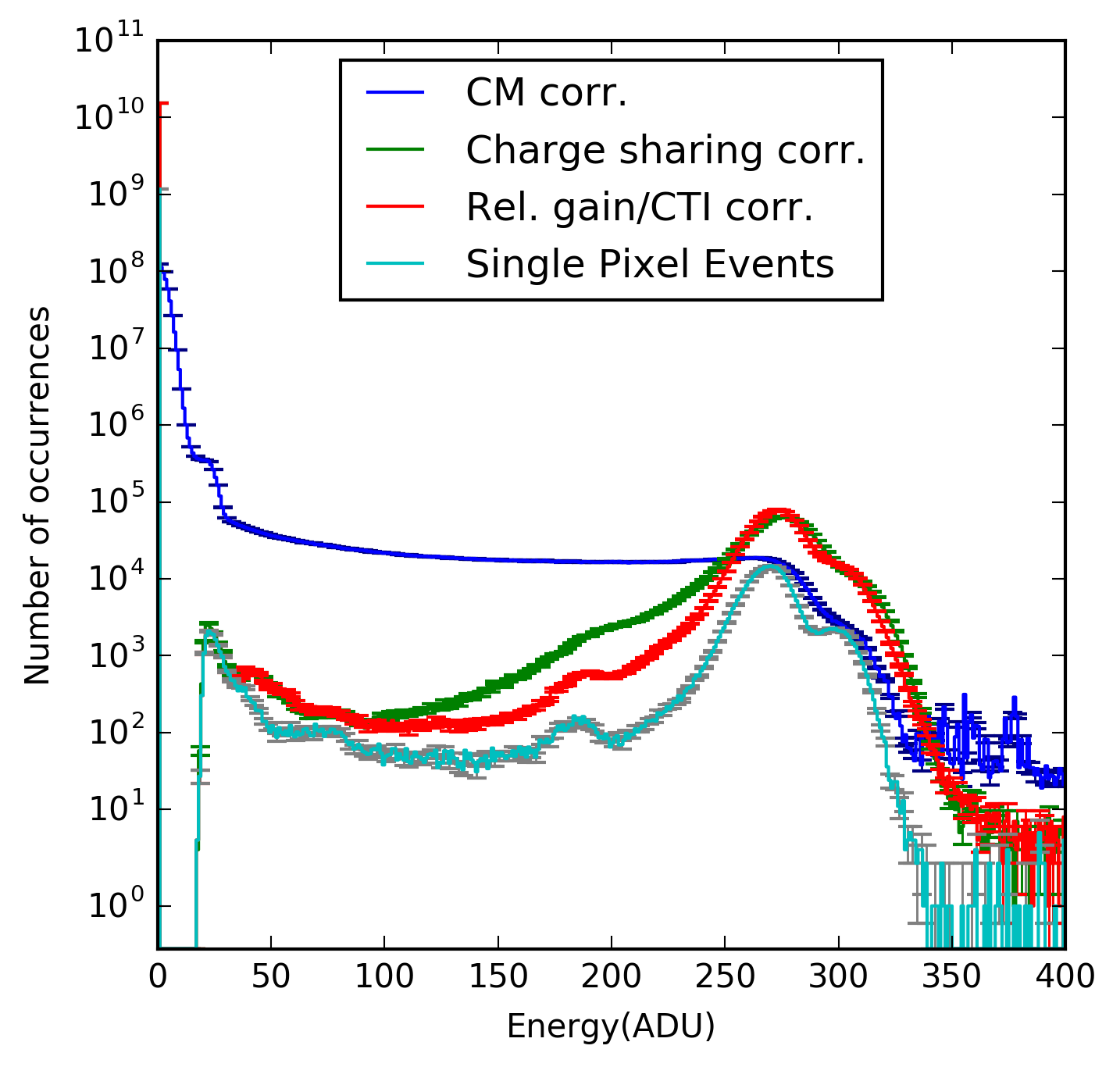}\vfill
\caption{Left: Combined relative gain and CTI map for the
	FastCCD. Right: $^{55}$Fe spectrum with Mn-K$_{\alpha}$ and
	Mn-K$_{\beta}$ peaks measured with the FastCCD. Improvement of
	the line width by applying data corrections is
	observable. The histogram labelled as "CM corr." corresponds to common-mode corrected data.
	Performing the corrections allows to reduce peak
	width almost to the level comparable to spectrum measured with
	a single pixel. }\label{fig:RG-CTI}
\end{figure}

Performing a flat-field analysis on a CCD detector allows
investigation of effects such as charge transfer inefficiency (CTI)
and varying relative gain due to different characteristics of the
readout amplifiers. CTI is a measure of the charge transfer losses given by 
\begin{align}
\label{eq:CTI}
CTI = 1 - \frac{Q_{n+1}}{Q_n},
\end{align}
where $Q_n$ is the charge in a pixel after n transfer cycles.
The per-column CTI of the FastCCD is shown in
Figure~\ref{fig:columnCTI}, where the top plot represents the
lower hemisphere (LH) and the bottom plot the upper hemisphere (UH) of the
FastCCD. 
Figure~\ref{fig:columnRG} shows column-wise relative gain with a
variation of 4.6\% for the LH (top plot)  and 4.9\% for the UH (bottom plot). The combination of CTI and relative gain
effects of the FastCCD can be seen in the left plot of Figure~\ref{fig:RG-CTI}, where the contribution of CTI and relative gain was calculated
for each pixel according to its distance from the readout amplifier and thus the number of shifts.

\begin{figure}[htbp]
\centering
\includegraphics[width=.47\linewidth, valign=t, trim=0 9 0 7, clip=True]{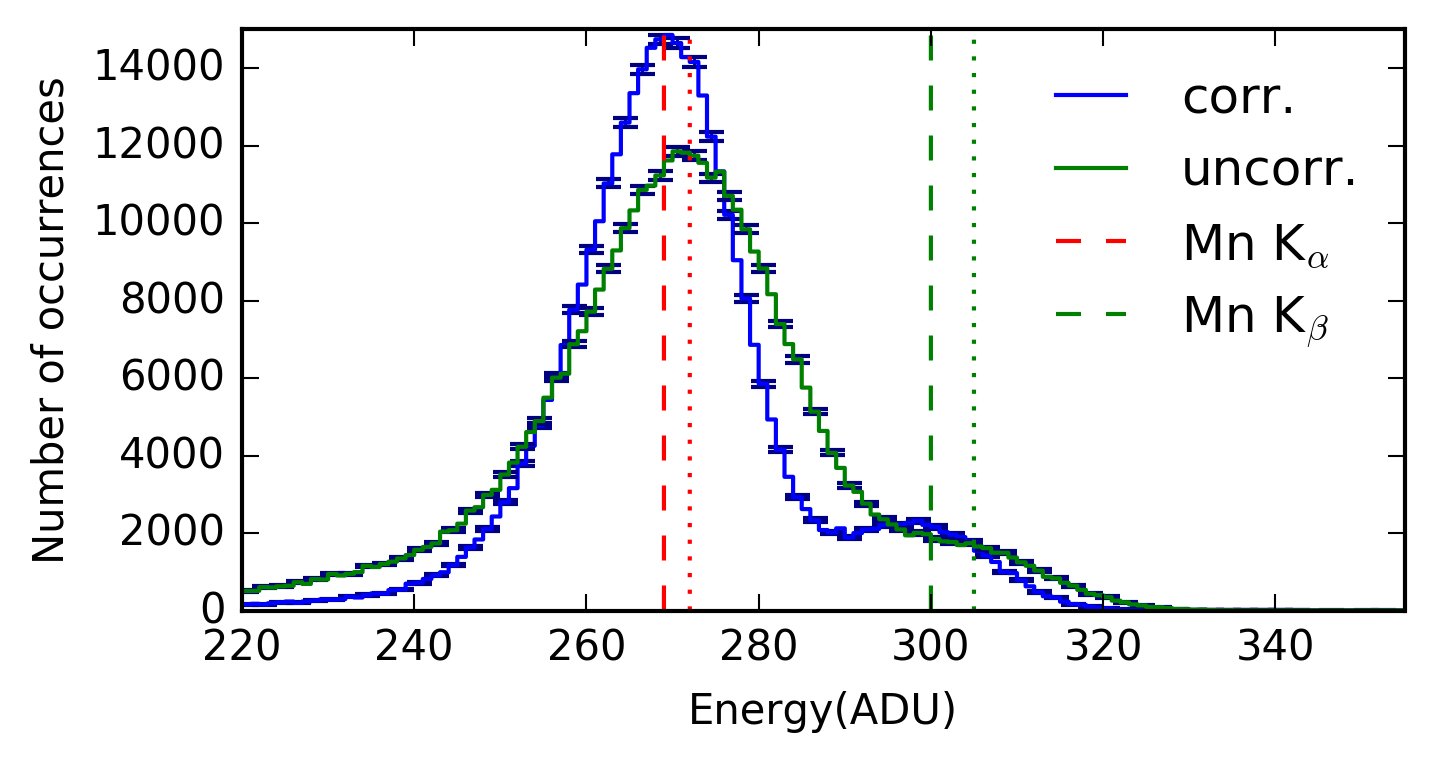}
\qquad
\includegraphics[width=.47\linewidth, valign=t, trim=0 9 0 9, clip=True]{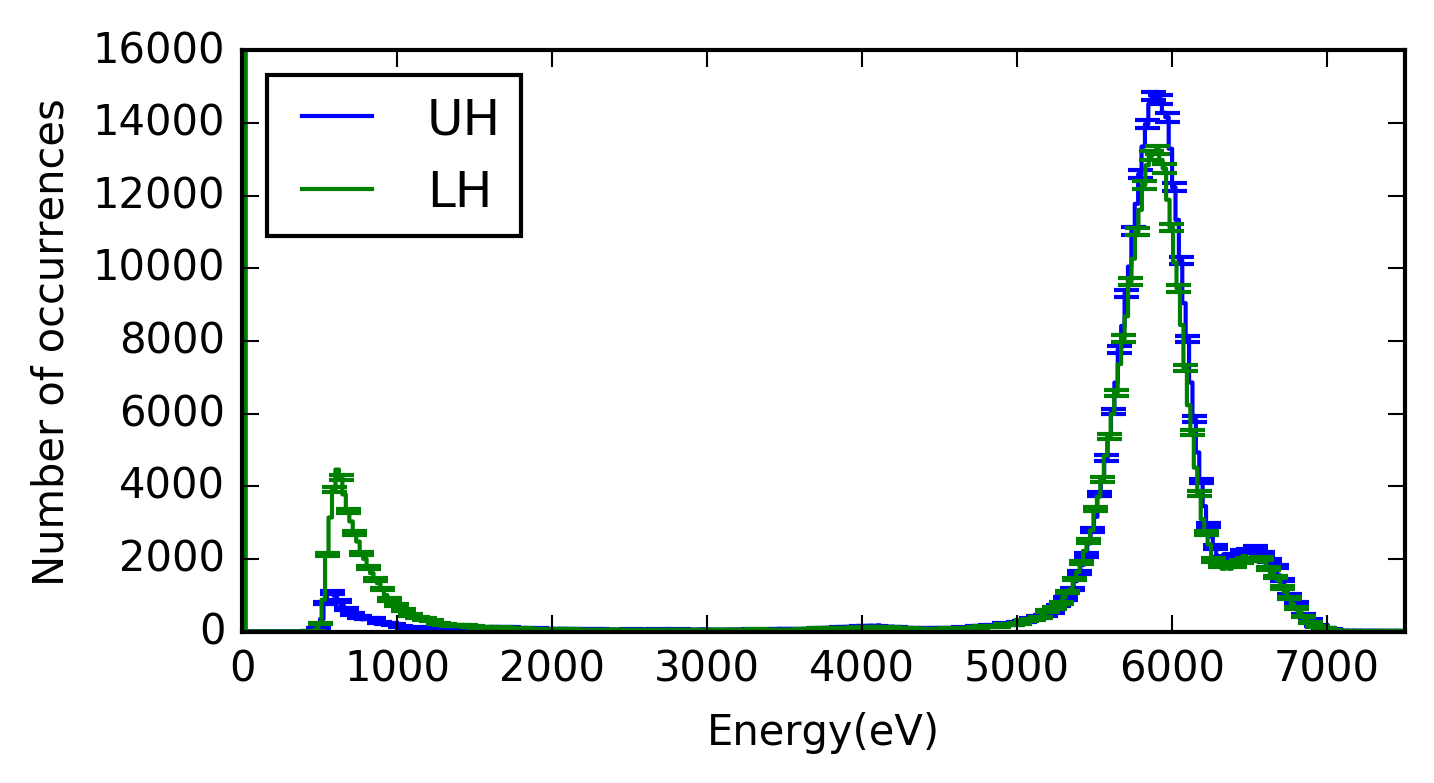}
\caption{Left: Comparison between CTI/gain corrected and
	uncorrected Mn-K$_{\alpha}$ and K$_{\beta}$ spectra taken with the FastCCD. The
	influence of applying CTI and relative gain correction
	results in a shift and narrowing of both peaks. Dotted red (Mn K$_{\alpha}$) and green (Mn K$_{\beta}$) lines show position of the peaks before corrections, whereas the dashed ones point to the peak positions after corrections. Right: Energy calibrated spectrum.}\label{fig:spectra}
\end{figure}

Due to the small pixel size of the FastCCD and ePix detectors, the
majority of charge is split across neighbouring pixels during the
charge collection process, affecting e.g. the energy resolution of the
device if this effect is not taken into account. By summing up the
contribution of all pixels the initial charge is distributed to, this
effect can be corrected for. As can be seen from the right plot of
Figure~\ref{fig:RG-CTI} and the left plot of Figure~\ref{fig:spectra} the
resulting energy resolution improves significantly from $\approx$
571$\,$eV to $\approx$ 426$\,$eV after correction (FWHM,
Mn-K$_{\alpha}$ peak), leading to a peak separation of 5.6$\,\sigma$
at 1$\,$keV deduced by extrapolation of the values corresponding to Mn K$_{\alpha}$ and Mn K$_{\beta}$. Performing the energy calibration, the resulting
$^{55}$Fe spectra can be transformed to energy units as is shown in
Figure~\ref{fig:spectra} (right). The detector performance as reported meets
the requirements set by the scientific instrument SCS.

\subsection{ePix100a Evaluation}

The ePix100a is an active pixel detector (with sensitive area of 704$\times$768 pixels), from a design-perspective the CCD-specific algorithms for gain evaluation (CTI + per-column relative gain) are thus not applicable. Rather a per-pixel evaluation, as described in~\cite{blaj2017robust} is the appropriate approach. However, the gain map deduced in the reference shows structuring which might be reproduced by combining a CTI-like effect with a per-column gain. As these algorithms were readily available from the FastCCD characterization, an evaluation using these techniques was attempted; the per-pixel approach is considered as a reference.

\begin{figure}[htbp]
\centering 
\includegraphics[width=.47\textwidth, trim=0 10 0 10, clip=True]{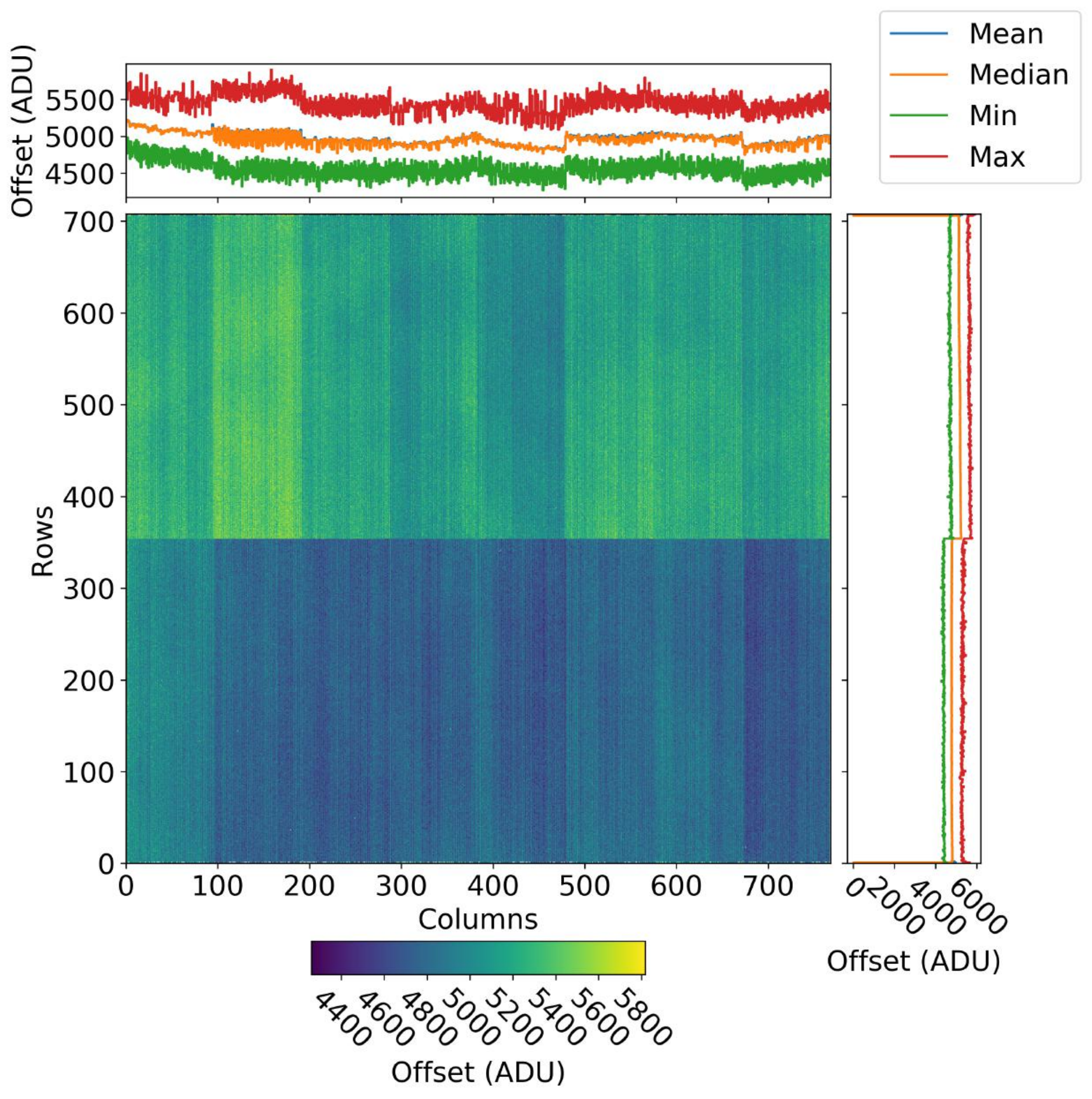}
\qquad
\includegraphics[width=.47\textwidth, trim=0 10 0 10, clip=True]{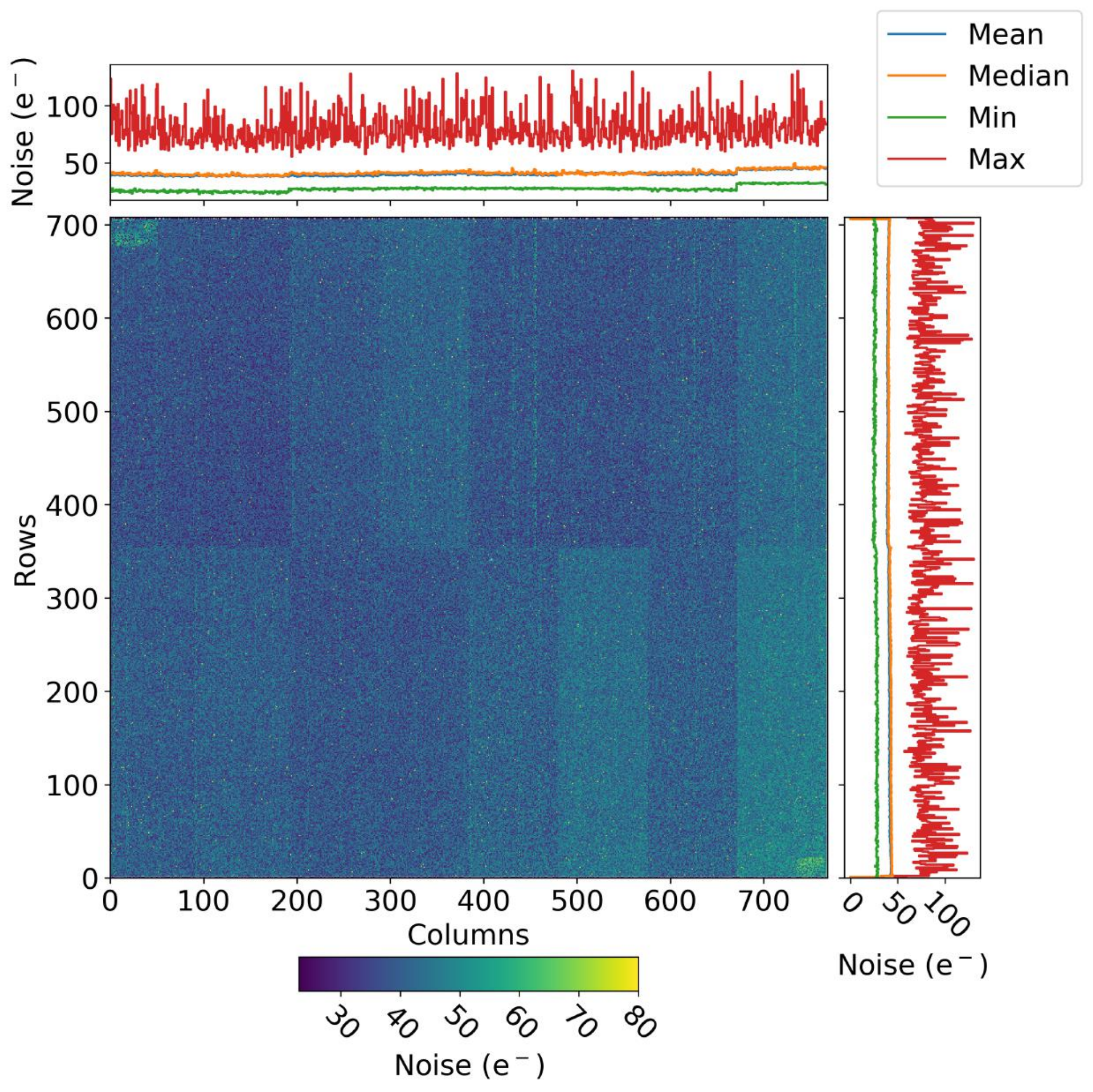}
\caption{\label{fig:darkepix}Offset map (left plot, in analog-to-digital units, ADU) and noise map (right plot) of the ePix100a taken at 10$\,^{\circ}$C with 50$\,\mu$s integration time. The side panels represent profiles along columns and rows.}
\end{figure}

 Figure~\ref{fig:darkepix} shows the resulting dark maps of
the ePix100a measured at a temperature of 10$\,^{\circ}$C with
50$\,\mu$s integration time. A strip pattern can be observed in the
offset map, due to the readout structure of the camera. The four tile
per ASIC like structure visible in the offset map originates in
multiplexing columns to four banks. The resulting RMS noise of the
ePix100a was found to be (40.5$\pm$7.2$)\,\mathrm{e^-}$, which is
within the specifications defined by scientific instruments HED and
MID.

\begin{figure}[htbp]
\centering 
\includegraphics[width=.31\textheight, valign=t, trim=0 5 0 5, clip=True]{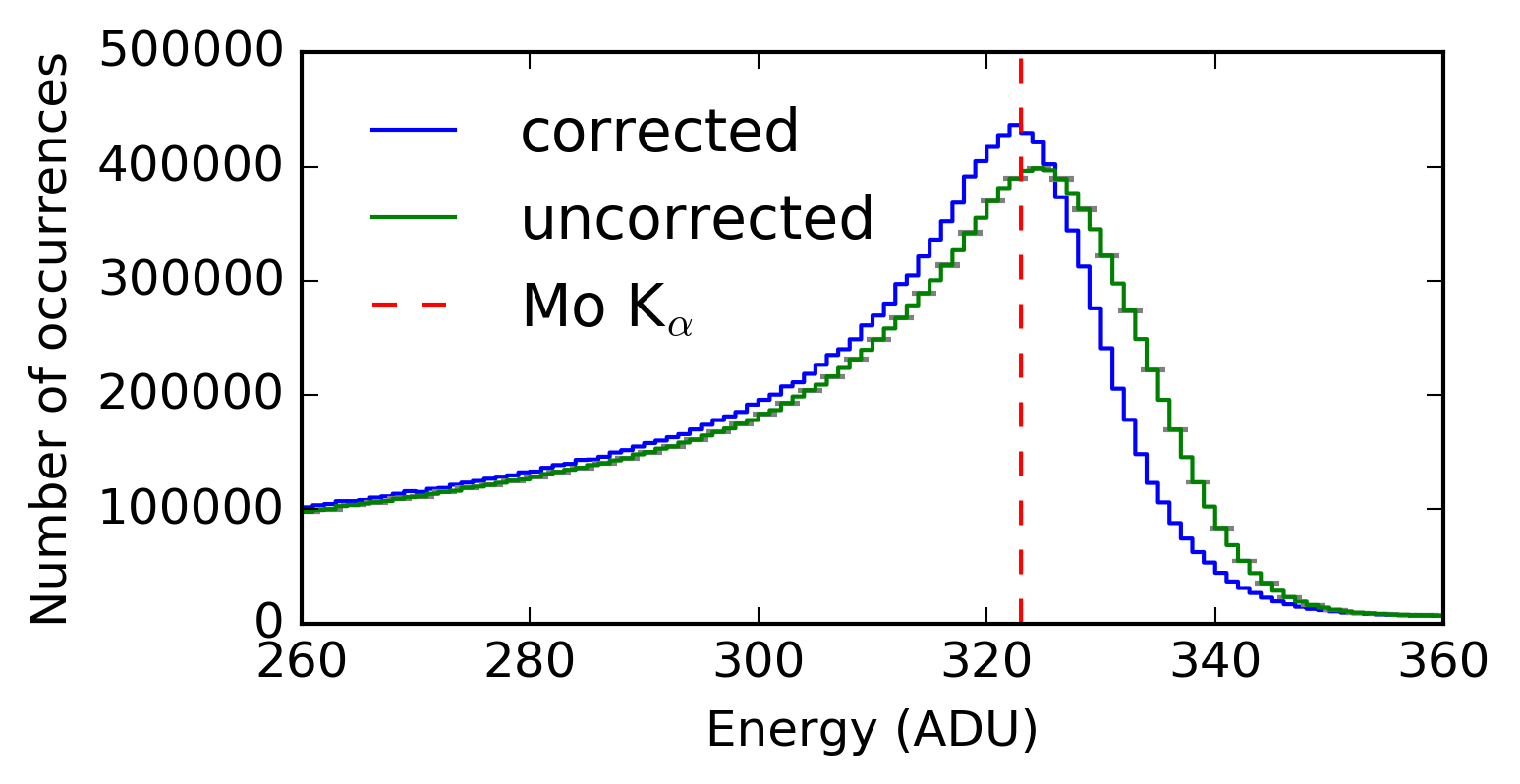}
\qquad
\includegraphics[width=.3\textheight, valign=t, trim=0 0 0 1, clip=True]{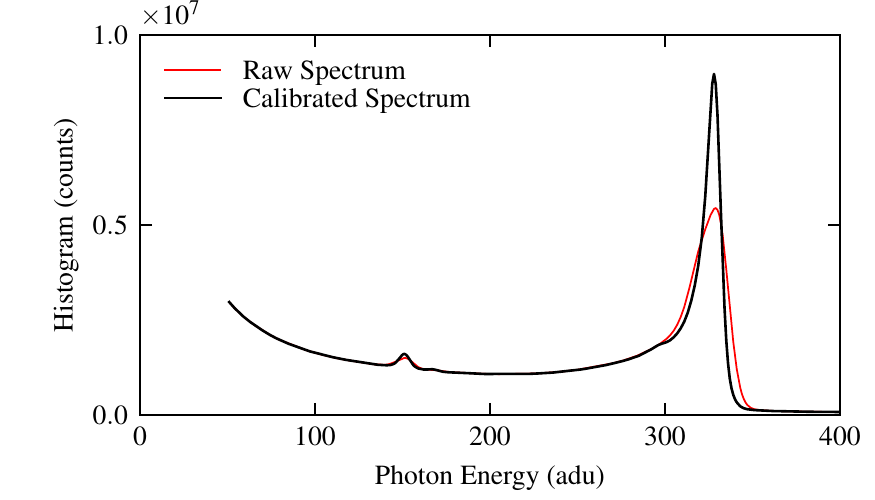}

\caption{\label{fig:moepix}Left: Comparison between the raw Mo K$_{\alpha}$ peak (green line) and the same peak after applying corresponding column-wise gain corrections (blue line). Right: The result of per-pixel gain calibration with 17.5$\,$keV (Mo K$_{\alpha}$) X-ray photons (Reproduced with permission from~\cite{blaj2017robust}), showing the improvements in energy resolution from $\approx$ 900$\,$eV (red line) to 590$\,$eV FWHM (black line) after per-pixel gain calibration.}
\end{figure}

Figure~\ref{fig:moepix} shows a Mo-K$_{\alpha}$ spectrum corrected
for per-column gain variations (left plot), and one corrected using the reference per-pixel correction from~\cite{blaj2017robust}. 
Using per-pixel gain corrections an energy resolution of $\approx$ 590$\,$eV was obtained: a $41\%$ improvement. 
Using the phenomenologically motivated CCD-like corrections, the energy resolution could be improved by about 30\%, to $\approx$ 760$\,$eV (FWHM). For an $^{55}$Fe source an energy resolution of $\approx$ 450$\,$eV (FWHM) at 5.9$\,$keV was determined using per-pixel corrections~\cite{blaj2018performance}. So far, $^{55}$Fe data taken at EuXFEL has only used an uncooled detector and thus no readily comparable values exist. Note that the energy resolution for a synchronous source, such as a FEL, is expected to improve further, as events outside the detector integration window are avoided.

\section{Conclusions}
\label{sec:out}

Our results show, that the performance of the FastCCD and ePix100a detectors is within the required specifications; for ePix100a even without gain corrections. For the FastCCD proper gain correction improves the energy resolution by 25\%, or to 426$\,$eV at 5.9$\,$keV. 
The corrections are necessary to reach >5$\,\sigma$ peak separation at 1$\,$keV. \par 
Applying a CCD-like characterization to the active pixel ePix100a detector shows 
that improvements in energy resolution are possible, 
as its per-pixel gain characteristics can phenomenologically be approximated 
by combining CTI-like and per-column relative gain effects. 
However, the detector design-driven per-pixel approach presented in \cite{blaj2017robust} outperforms the phenomenologically driven approach by $\approx 25\,\%$ in terms of obtainable energy resolution, with low requirements on photon counts. Hence, we plan on integrating the low-pixel-statistics gain evaluation of \cite{blaj2017robust} into EuXFEL calibration tool-chain.

\acknowledgments

We would like to acknowledge the support of U. B\"{o}senberg and A. Parenti in detector integration at the European XFEL. This work was supported in part by the Department of Energy contract DE-AC02-76SF00515.

\bibliography{references/referenceIW}
\bibliographystyle{unsrt}
\nocite{*}

\end{document}